\RequirePackage{lineno}
\documentclass[aps,prb,showpacs, groupedaddress,twocolumn,superscriptaddress]{revtex4}
\usepackage{amsmath,graphicx,latexsym,times,color}
\usepackage{setspace}
\usepackage{hyperref}
\usepackage{array}
\usepackage{titlesec}
\usepackage{physics}

\begin{document}

\title{Tuning spin filtering by anchoring groups in benzene derivative molecular junctions}

\author{Dongzhe Li}
\email{dongzhe.li@uni-konstanz.de}
\affiliation{Department of Physics, University of Konstanz, 78457 Konstanz, Germany}

\author{Yannick J. Dappe}
\affiliation{Service de Physique de l'Etat Condens\'e (SPEC), CEA, CNRS, Universit\'e Paris-Saclay, CEA Saclay 91191 Gif-sur-Yvette Cedex, France}

\author{Alexander Smogunov}
\affiliation{Service de Physique de l'Etat Condens\'e (SPEC), CEA, CNRS, Universit\'e Paris-Saclay, CEA Saclay 91191 Gif-sur-Yvette Cedex, France}

\date{\today}

\begin{abstract}
    One of the important issues of molecular spintronics is the control and manipulation of charge transport and, in particular, its spin polarization through single-molecule junctions. Using $ab$ $initio$ calculations, we explore spin-polarized electron transport across single benzene derivatives attached with six different anchoring groups (S, CH$_3$S, COOH, CNH$_2$NH, NC and NO$_2$) to Ni(111) electrodes. We find that molecule-electrode coupling, conductance and spin polarization (SP) of electric current can be modified significantly by anchoring groups. In particular, a high spin polarization (SP $>$ 80\%) and a giant magnetoresistance (MR $>$ 140\%) can be achieved for NO$_2$ terminations and, more interestingly, SP can be further enhanced (up to 90\%) by a small voltage. The S and CH$_3$S systems, on the contrary, exhibit rather low SP while intermediate values are found for COOH and CNH$_2$NH groups. The results are analyzed in detail and explained by orbital symmetry arguments, hybridization and spatial localization of frontier molecular orbitals. We hope that our comparative and systematic studies will provide valuable quantitative information for future experimental measurements on that kind of systems and will be useful for designing high-performance spintronics devices.
    
\end{abstract}

\renewcommand{\vec}[1]{\mathbf{#1}}

\maketitle

\section{ Introduction }
\label{Intro}

Molecular (organic) spintronics \cite{sanvito2011} is a rapidly developing field of research, aiming at the manipulation of both electron charge and spin in molecular-based devices, taking advantage of large spin relaxation length across purely organic molecules due to small spin-orbit interactions. The one of most fundamental and crucial properties here is the spin polarization ($\text{SP}$) of the current by ferromagnet/organic interface \cite{Atodiresei-2010,Barraud-2010, Requist-2016}, which can be defined as $\text{SP}=(G_{\downarrow}-G_{\uparrow})/(G_{\uparrow}+G_{\downarrow}) \times 100\%$, where $G_{\uparrow}$ and $G_{\downarrow}$ are spin up (majority) and spin down (minority) conductance, respectively. Recently, it has been shown that the SP in single-molecule junctions can be tunned by a mechanical strain \cite{tang2015strain}, an orbital symmetry consideration \cite{Alex_2015}, and spin-dependent quantum interference effect \cite{Valli2018,Dongzhe-QI2019,valli2018interplay} etc. Understanding physical and chemical mechanisms involved in spin injection at the hybrid interfaces for further design of possible molecular-based devices with large SP and high conductance is one of the most important issues in this field.

The anchoring groups (also known as ``linkers'') placed at extremities of the molecule are responsible for establishing a stable mechanical contact and efficient molecule/metal electronic coupling. For this reason, thiol ($-$SH) \cite{Chen-2006} and thiolate ($-$S) \cite{Ke-2004,Venkataraman-2006} have become the most widely used anchoring groups due to strong covalent gold-sulfur bonding. However, it was argued extensively that the conductance of thiol-based molecular junctions depends strongly on the binding geometry \cite{li2008charge}. Therefore, many theoretical and experimental efforts were made to explore various anchoring groups such as methythiol ($-$CH$_3$S) \cite{Park-2007,Meisner-2011}, carboxyl-acids ($-$COOH) \cite{Chen-2006,Ahn-2012}, amidine ($-$NH$_2$) \cite{Venkataraman-2006,Farzadi2018}, isonitrile ($-$NC) \cite{Xue-2004,Veronika2015}, nitrile ($-$N) \cite{kaminski2016tuning}, nitro ($-$NO$_2$) \cite{Venkataraman-2007,Tsukamoto-2012}, etc. These investigations ended up with two general conclusions: first, the chemical nature of anchoring groups strongly affects the energy level alignment of molecular frontier orbitals with respect to the metal Fermi level; second, the degree of hybridization between the molecule and metal changes dramatically with anchoring groups. 

For 3$d$ ferromagnetic materials, the $s$ band is almost non spin-polarized while the 3$d$ bands are spin-split due to the exchange interactions. As a result, the density of states (DOS) of 3$d$ ferromagnetic materials (such as Fe, Co, or Ni) shows a spin polarization of about 30$\sim$40$\%$ at the Fermi energy. Therefore, when organic molecules are contacted with 3$d$ metals, a selective hybridization occurs at the molecule-metal interface for spin up and down channels. For example, due to large $\pi-d$ hybridization at the ferromagnetic metal/organic interfaces, high spin polarization \cite{Atodiresei-2010,Dongzhe-2016}, controllable ferro- or antiferromagnetic interlayer exchange coupling \cite{Arnoux2019}, giant magnetoresistance \cite{cakir-2014} and enhanced perpendicular magnetic anisotropy \cite{Bairagi-2015} were reported (a detailed discussion on molecular spinterface can be found in Ref. \cite{Cinchetti-2017}).

Previous studies were mainly focused on the effect of anchoring groups on charge transport properties with nonmagnetic electrodes (Au, Ag and Cu, etc). The investigation of spin polarization via various anchoring groups at organic spinterface was pointed out recently \cite{ZHANG201760,Qiu2018}. In this work, based on spin-polarized $ab$ $initio$ transport calculations, we present a comparative and systematic study of the impact of anchoring groups on spin-dependent transport with ferromagnetic electrodes. More specifically, we have chosen a benzene as a core structure and have studied the spin-dependent transport for a series of Ni(111)/X$-$(C$_6$H$_4$)$-$Y/Ni(111) junctions, where terminations X, Y could be S, CH$_{3}$S, COOH, CNH$_{2}$NH, NC or NO$_2$ groups, as demonstrated in Fig. \ref{structure}. These systems are chosen because Ni-based spin valves were firstly proposed by Emberly $et$ $al$ \cite{emberly2002molecular} as a prototypical molecular spintronic system, and later on were studied extensively by several theory groups based on $ab$ $initio$ methods \cite{Waldron-2006, rocha2005towards} and also successfully created by mechanically controlled break junction (MCBJ) experiment \cite{rakhmilevitch2016enhanced}. Importantly, as will be discussed later, all these molecules have frontier orbitals which, by symmetry, do not overlap with the Ni electrode's $s$ states. Due to symmetry arguments proposed by us recently \cite{Alex_2015,Dongzhe-2016-spin}, all the junctions are therefore expected to display rather high SP of conductance which is indeed confirmed by our calculations. We find, moreover, that among all considered molecules the one with NO$_2$ terminations, M6 in Fig. \ref{structure}, presents very high values of SP and of total conductance at the same time. In addition, high spin filtering in M6 ($-$NO$_2$) is accompanied by huge magnetoresistance (MR) ratio (about 140\%) which measures the change in resistance (or conductance) between parallel and antiparallel magnetic configurations of two ferromagnetic electrodes. These findings make therefore M6 molecule the most promising candidate for possible spintronics applications. 

The paper is organized as follows. In Sec. \ref {method}, we present computational methods and models used in this work. In Sec. \ref {Results and discussion}, the electronic structure and transport properties of benzene derivative molecules with different anchoring groups in equilibrium will be presented. Then, we will present the non-equilibrium transport phenomena with a particular focus on their SP and MR. Finally, the conclusion will be drawn in Sec. \ref{conclusions}.

\section{ Calculation methods and models}
\label{method}

\begin{figure*}[htbp]
	\centering
	\includegraphics[scale=0.55]{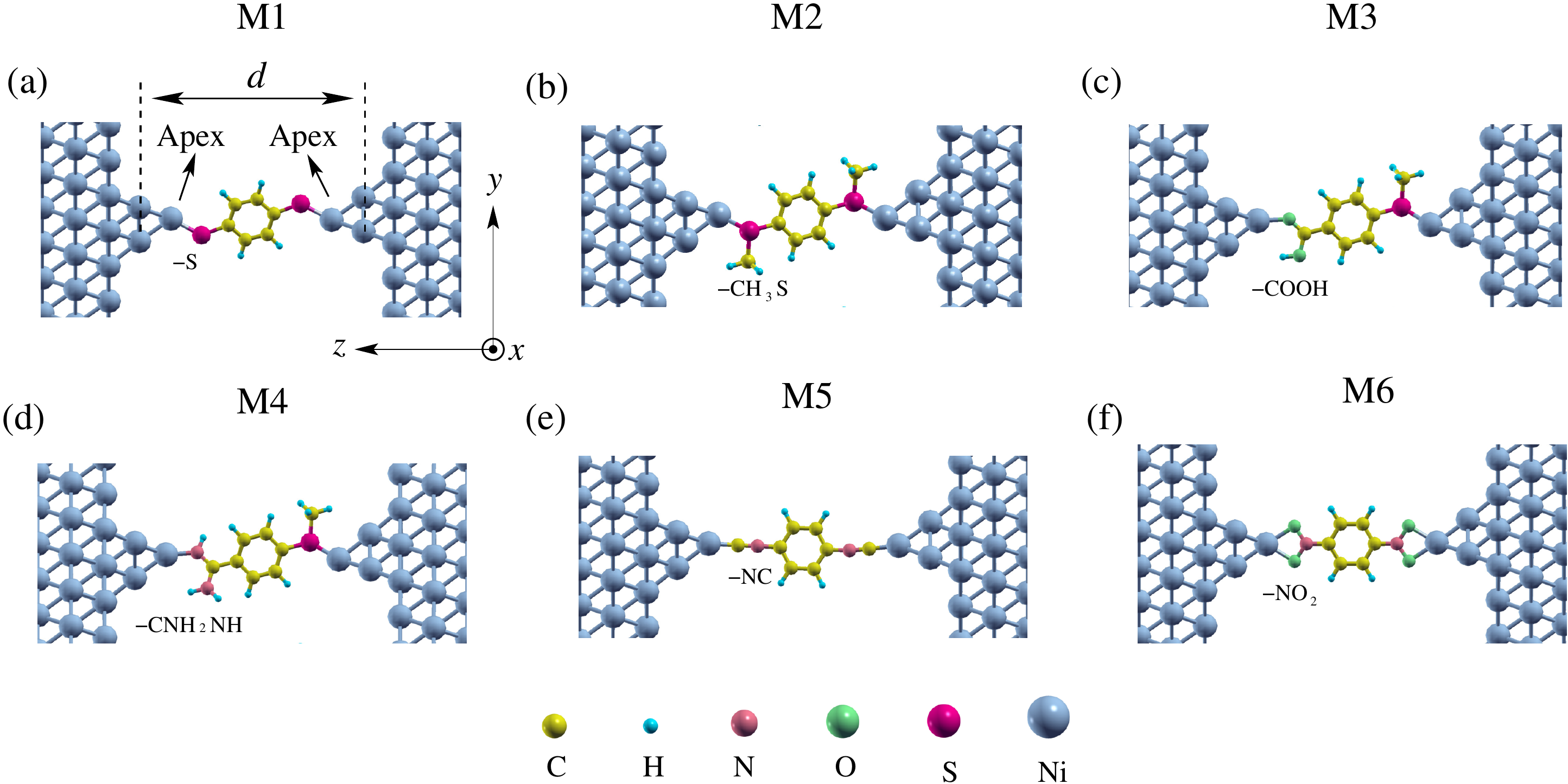}
	\caption{\label{structure}
		Schematic representation of optimized atomic structures for benzene-based molecules with different anchoring groups connecting two Ni(111) electrodes. The investigated six molecules in this work are: (a) 1,4-benzenedithiolate (M1), (b) benzene, 1,4-bis (methylthio) (M2), (c) 4-(methylthio) benzonic acid (M3), (d) benzene, 1,4-amine and methylthio (M4), (e) 1,4-phenylene diisocyanide (M5) and (f) 1,4-dinitrobenzene (M6). Note that M3 and M4 are asymmetric molecular junctions with different anchoring groups on the left and right sides. $Z$ is the charge transport direction which is parallel to the junction axis.
	}
\end{figure*}

The geometry optimization of molecular junctions was carried out using plane waves  $\textsc{Quantum-Espresso}$ (QE) package \cite{Giannozzi2009} within the density functional theory (DFT). We used Perdew–Burke–Ernzerhof \cite{PBE-1996} exchange-correlation functionals and ultrasoft pseudopotentials (PP) to describe electron-ion interactions. Plane-wave energy cutoffs of 30 and 300 Ry were used for wave functions and charge density, respectively. Molecular junctions were described in a supercell containing a molecule and two four-atom Ni pyramids attached to both sides to Ni(111) slabs with 4 $\times$ 4 periodicity in the $XY$ plane (16 atoms per layer) containing six layers on each side as shown in Fig. \ref{structure}. During ionic relaxation, three bottom layers on both sides were kept fixed at their bulk positions while a molecule and other Ni layers were allowed to relax until atomic forces were less than 10$^{-3}$ Ry/Bohr. Relaxation was performed using (2 $\times$ 2 $\times$ 1) $k$-point mesh.

After the atomic relaxation, $ab$ $initio$ spin-polarized electronic transport properties for different molecular junctions were evaluated using the $\textsc{Transiesta}$ code \cite{Brandbyge-2002} which employs a non-equilibrium Green's function (NEGF) formalism combined with DFT. We used Troullier-Martins norm-conserving pseudopotentials \cite{Troullier-1993}, PBE functional and an energy cutoff for the real-space mesh of 250 Ry. Valence electron wavefunctions were expanded in a basis of local orbitals in $\textsc{siesta}$ \cite{soler2002siesta}. A double $\zeta$ plus polarization (DZP) basis set with an energy shift of 50 meV was used, which resulted, as we have checked, in a good agreement with QE results for both magnetic properties and energy alignments (see Fig. \ref{Sup-fig1} in Appendix A). The convergence tolerance for self-consistent loop was set to 10$^{-4}$ eV and the Brillouin zone was sampled by 6 $\times$ 6 $\times$ 1 $k$-point mesh.

\begin{center}
	\begin{table*}[htbp]
		
		\scalebox{1.0}{
			\begin{tabular}{cccccccccc}
				\hline\hline
				\multicolumn{1}{c}{$$} & \multicolumn{1}{c} {$d$ ($\text{\AA}$)} & \multicolumn{1}{c} {$d_{\text{Ni-X}}$ ($\text{\AA}$)} & \multicolumn{1}{c} {$d_{\text{Ni-Y}}$ ($\text{\AA}$)} & \multicolumn{1}{c} {$M_{s}$ ($\mu_{\text{B}}$)} & \multicolumn{1}{c} {$G_{\uparrow}$ ($G_0=e^2/h$)} & \multicolumn{1}{c} {$G_{\downarrow}$ ($G_0=e^2/h$)} & \multicolumn{1}{c} {$G_{\text{tot}}$ ($G_0=e^2/h$)} & \multicolumn{1}{c} {$\text{SP}$ (\%)}\\ \hline
				M1 & 13.78 &  2.14 & 2.14  & 0.28 & 1.48 $\times$ 10$^{-1}$ & 7.15 $\times$ 10$^{-2}$ & 2.20 $\times$ 10$^{-1}$ & -34.99 \\ 
				M2 & 13.45 & 2.16 & 2.16 &  -0.02 & 9.81 $\times$ 10$^{-3}$ & 7.77 $\times$ 10$^{-3}$ & 1.76 $\times$ 10$^{-2}$ & -11.50 \\ 
				M3 & 14.03 & 2.11 & 1.89  & -0.07 & 1.73 $\times$ 10$^{-2}$ & 8.80 $\times$ 10$^{-2}$ & 9.77 $\times$ 10$^{-2}$ & 64.49 \\ 
				M4 & 14.24 & 2.12 & 1.89  & -0.04 & 1.51 $\times$ 10$^{-2}$ & 4.10 $\times$ 10$^{-2}$ & 5.62 $\times$ 10$^{-2}$ & 46.00 \\  
				M5 & 15.25 & 1.78 & 1.78  & -0.14 & 1.67 $\times$ 10$^{-2}$ & 2.78 $\times$ 10$^{-1}$ & 2.50 $\times$ 10$^{-1}$ & 87.46 \\ 
				M6 & 14.24 & 2.04 & 2.04  & 0.31 & 3.54 $\times$ 10$^{-2}$ & 3.21 $\times$ 10$^{-1}$ & 3.56 $\times$ 10$^{-1}$ & 80.13  \\  \hline\hline
				\end{tabular}}
			\caption{Optimized junction distances, induced molecular spin moment ($M_{s}$) (calculated from spin-polarized molecular DOS integrated up to the Fermi energy as shown in Fig. \ref{Sup-fig2}), spin-resolved and total conductances (in the unit of $G_0=e^2/h$ which is conductance quantum per spin) in parallel spin configuration and its spin polarization.} \label{table-1}
			\end{table*}
			\end{center}

Spin-resolved (denoted by spin index $\sigma= \uparrow, \downarrow$) transmission function, depending on energy $E$ and applied bias $V_{\text{b}}$, is given by:

\begin{equation}
T_{\sigma}=\mathrm{Tr}[\Gamma_{L,\sigma}G_{\sigma}^r\Gamma_{R,\sigma}G_{\sigma}^a],
\end{equation}

where all matrices depend also on $E$ and $V_{\text{b}}$ and have dimension of the scattering region (or extended molecule) including the molecule itself and some parts of left and right electrodes (where screening takes place). $G_{\sigma}^{r/a}$ are the retarded/advanced Green's functions:

\begin{equation}
G_{\sigma}^{r/a}=[(E\pm i\eta)S - H_{\sigma}^{C} - \Sigma_{L,\sigma}^{r/a} - \Sigma_{R,\sigma}^{r/a}]^{-1}
\end{equation}

with $\eta$ is an infinitesimal positive number, $S$ is the overlap matrix, $H_{\sigma}^C$ is the Hamiltonian matrix for the scattering region and $\Sigma_{L/R,\sigma}^{r/a}$ are retarded or advanced self-energies due to left/right electrodes. Coupling matrices $\Gamma_{L/R,\sigma}$ are evaluated from corresponding self-energies as $\Gamma_{L/R,\sigma} = i (\Sigma_{L/R,\sigma}^r-\Sigma_{L/R,\sigma}^a)$.

Finally, the spin-dependent charge current is obtained from the Landauer-B\"{u}ttiker formula:

\begin{equation}\label{Landauer}
I_{\sigma} (V_{\text{b}}) = \frac{e}{h}\int_{-\infty}^{+\infty}dE[f(E, \mu_L) - f(E, \mu_R)]T_{\sigma}(E, V_{\text{b}}),
\end{equation}

where $f(E, \mu_{L/R})$ are Fermi-Dirac distribution functions, and $\mu_{L/R}$ are electrochemical potential of left/right electrodes.

\section{Results and discussion}
\label{Results and discussion}

The optimized geometries of molecular junctions are shown in Fig. \ref{structure}. We first performed atomic relaxations of a Ni/molecule interface to obtain an electrode-molecule separation and geometry, and then carried out full relaxations by attaching the second electrode at the previously calculated molecule-metal distance. Let us stress that relaxations have been performed starting from several possible initial configurations in order to find the minimum energy configuration. For example, the nitro ($-$NO$_2$)-terminated molecule prefers to bind through double Ni$-$O bonds on each side rather than with Ni$-$N or single Ni$-$O bonds as seen in Fig. \ref{structure} (f). Some important structural parameters, electronic and transport properties are summarized in Table~\ref{table-1}. Note that M3 and M4 are asymmetric junctions with two different linking groups on left and right sides while all the others are symmetric. Small induced (by Ni electrodes) spin moments on M1 and M6 were found to be positive (``ferromagnetic" molecule/Ni coupling) while for other molecules -- negative (``antiferromagnetic'' coupling). Spin-dependent conductance, $G_{\sigma}$, is given by the Landauer-B\"uttiker formula, $G_{\sigma}=G_0T_{\sigma}(E_F)$, where $G_0=e^2/h$ is the quantum conductance per spin ($e$ is the electron charge and $h$ is Plank's constant) and $T_{\sigma}(E_F)$ is the transmission function for spin $\sigma= \uparrow, \downarrow$ at the Fermi energy. The calculated total conductance (summed over spin up and down channels) of M1 at zero bias voltage was found to be about 0.22$G_0$ which is in agreement with previous calculations \cite{Lee-2014}. Additionally, the SP of M1 and M2 was found to be negative ($G_{\uparrow}$ $>$ $G_{\downarrow}$) while all the junctions have positive SP. Previous DFT calculations showed also a negative SP in Ni/BDT/Ni junction \cite{Waldron-2006}. Moreover, M5 and M6 exhibit large spin polarization of about 87$\%$ and 80$\%$ and high conductance of about 0.25$G_0$ and 0.35$G_0$, respectively, so they appear to be most attractive for possible future applications in spintronics devices.

\begin{figure*}[htbp]
	\centering
	\includegraphics[scale=0.60]{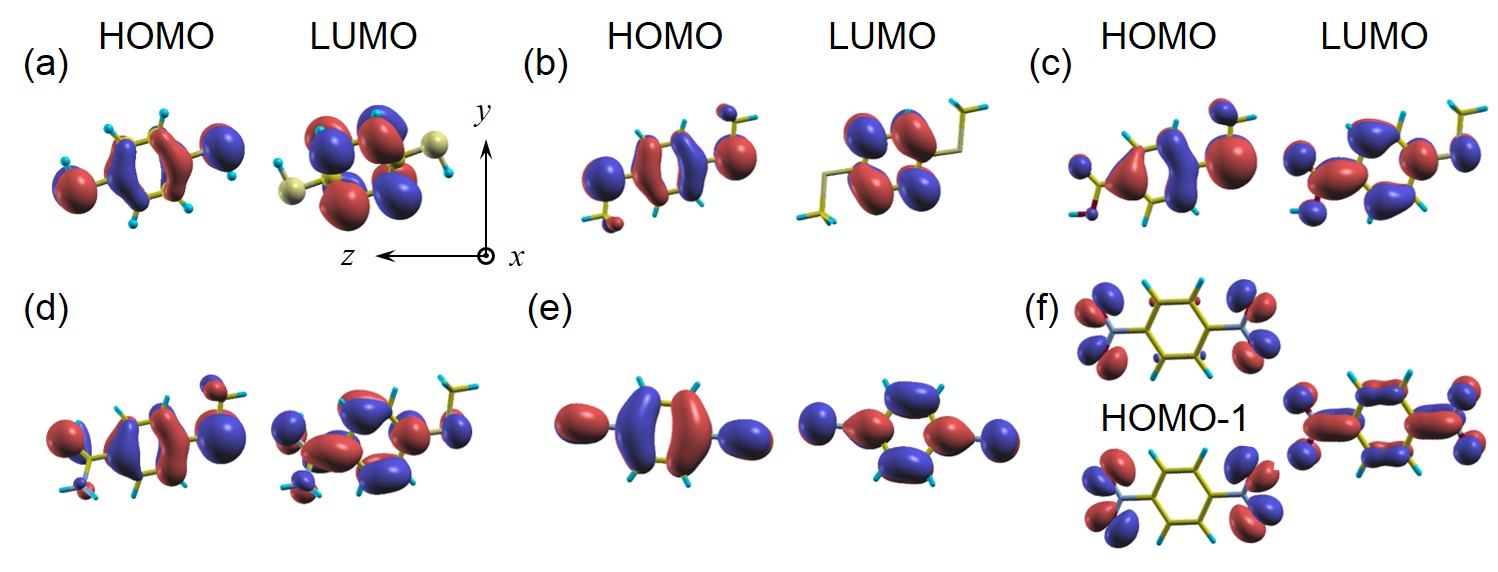}
	\caption{\label{homo-lumo}
		HOMO and LUMO molecular orbitals in the gas phase for the same molecules as in Fig.\ref{structure}. Isosurfaces of positive and negative isovalues are shown in red and blue, respectively. Note that all the orbitals are of $\pi$-type (odd with respect to the $YZ$ plane) except HOMO and HOMO-1 for M6 (which are even). The latter orbitals are nearly degenerate (split by about 0.1 eV) and represent bonding/anti-bonding combinations of end-group originated states.}
\end{figure*}

First, we plot in Fig. \ref{homo-lumo} highest occupied molecular orbitals (HOMO) as well as lowest unoccupied molecular orbitals (LUMO) for all the molecules in the gas phase. We found that all the orbitals are of $\pi$-type (odd with respect to the molecular $YZ$ plane) except HOMO and HOMO-1 for M6 case which is split by only about 0.1 eV and is of $\sigma$-type (even with respect to the $YZ$ plane). Interestingly, these M6 orbitals are both localized on $-$NO$_2$ anchoring groups forming a kind of bonding/anti-bonding states,
even or odd with respect to the transport direction $Z$. Since the molecules have relatively small tilting angles in the $YZ$ plane, all frontier $\pi$-orbitals have a rather small overlap with spin up $s$-states of Ni apex atoms (two Ni atoms which contact the molecule as indicated in Fig. \ref{structure}a)
while a strong coupling with spin down $d_{xz}$ states is expected. That should lead to rather strong SP of conductance due to orbital symmetry argument \cite{Alex_2015,Dongzhe-2016-spin}. This reasoning is also valid for HOMO and HOMO-1 orbitals of M6, though they are not of $\pi$-type.  They are still orthogonal to Ni apex $s$-states because of odd symmetry with respect to the $XZ$ plane and will only transmit spin down electrons injected by Ni apex $d_{yz}$ orbitals.     

\begin{figure*}[htbp]
	\centering
	\includegraphics[scale=0.46]{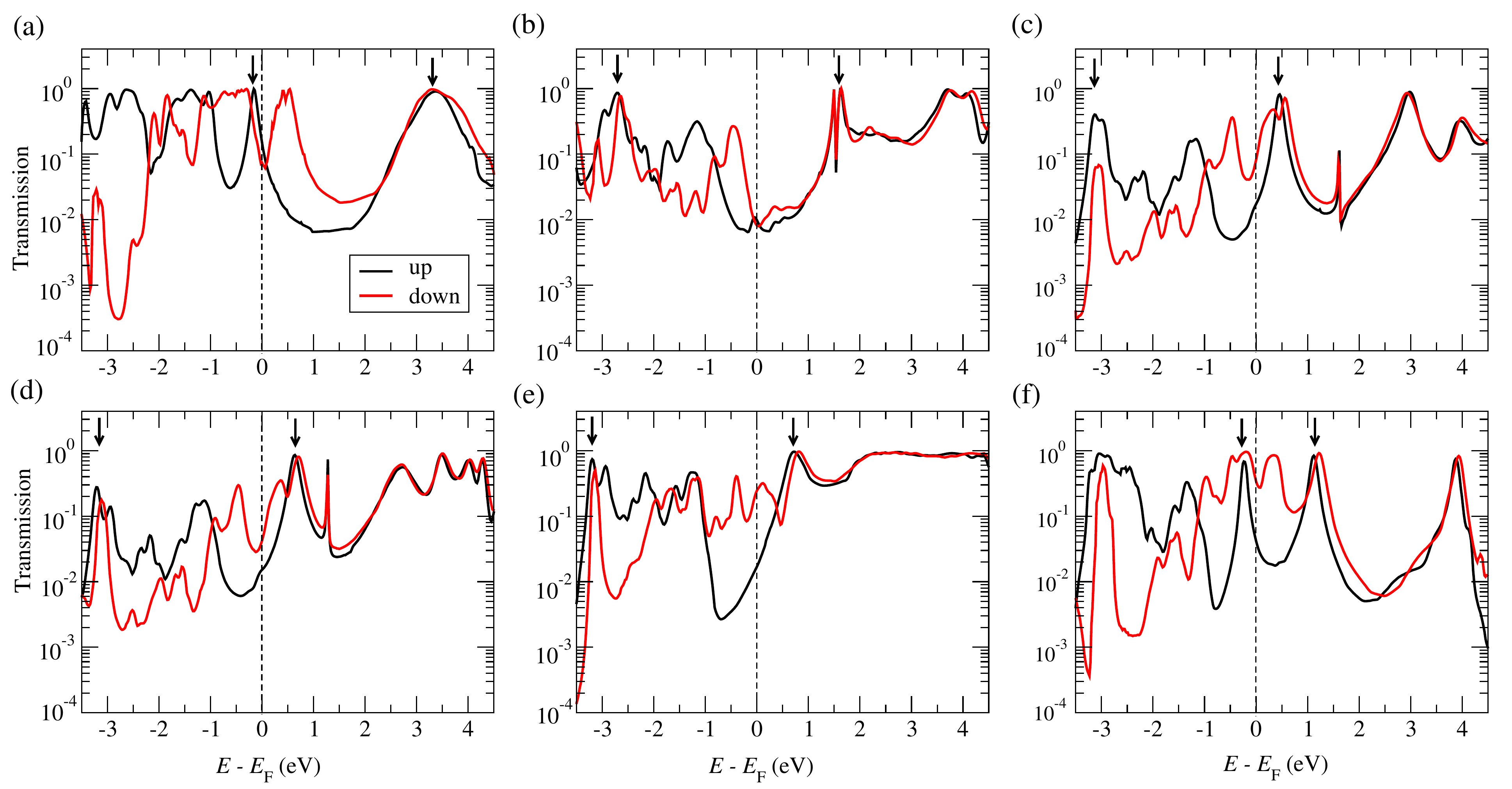}
	\caption{\label{T-E-curve}
		Spin-resolved zero-bias transmission functions (in logarithmic scale) with the parallel magnetic alignment of two Ni electrodes for (a) M1, (b) M2, (c) M3, (d) M4, (e) M5 and (f) M6 molecular junctions. Spin up and down channels are plotted by black and red lines, respectively. Note that the zero of energy is at the Fermi level. Positions of HOMO and LUMO for spin up is marked by black arrow.
	}
\end{figure*}

To better understand conductance as well its SP discussed above, we show in Fig. \ref{T-E-curve} spin-resolved transmission functions at zero bias in the parallel spin configuration of Ni electrodes. The black and red curves show the spin up and down transmissions, respectively.
First, we note that energy alignment of frontier molecular orbitals with respect to the Fermi energy, imposed by Ni electrodes, changes drastically with the anchoring groups. While for M1 and M6 we find that the transport is dominated by the HOMO (p-type current by holes), in the case of M2, M3, M4, and M5, the conduction takes place through the LUMO (n-type current by electrons). Second, the width of transmission features is attributed to the degree of molecule level hybridization with electrode states. Therefore, much more structured $T(E)$ with broader features are found for spin down due to extra $d_{\downarrow}$ states of Ni in the vicinity of the $E_F$. As a general feature, 
two peaks are often seen in spin down transmission: the first one at about $-0.5$ eV (see, e.g., Fig. \ref{T-E-curve}b-e) and another one at about $0.4$ eV (see, e.g., Fig. \ref{T-E-curve}a,c,d,f) which originate from $d^{\downarrow}_{x^2-y^2,xy}$ and $d^{\downarrow}_{zx,zy}$ states, respectively, of Ni apex atoms. Here, we mark the positions of HOMO and LUMO orbitals for spin up by black arrows. For spin down case, the HOMOs are much more delocalized in energy due to their strong hybridization with additional $d^{\downarrow})$ states of Ni apexes. For more details, see the molecular DOS in Fig. \ref{Sup-fig2} (Appendix B).

We first discuss the electron-donating anchoring groups such as M1 and M6. 
For M1, rather sharp spin up peak at about $-0.2$ eV originates from the HOMO weakly hybridizing with $s_{\uparrow}$ states of Ni apexes and with other orbitals of deeper Ni atoms. For the spin down, much broader feature in the transmission is observed at energies $-1.2$ eV$<E<1$ eV coming from the coupling of HOMO with $(s^{\downarrow}$ + $d^{\downarrow})$ states of Ni apexes. In particular, the peak at about $0.5$ eV is related to the offset of 
Ni $d^{\downarrow}_{xz}$ states appearing at about $E<1$ eV. Another peak in both spin channels at about 3.2 eV is attributed to the LUMO level. 
This result is in good agreement with previous $ab$ $initio$ calculations \cite{Lee-2014,Waldron-2006}. Interestingly, in the case of M6, compare to M1, the HOMO-derived spin up transmission peak is more sharp, resulting in significantly reduced spin up conductance and thus higher spin polarization. This can be explained by two reasons as follows. The first one is the orbital symmetry argument as mentioned before. During atomic relaxations, the planar configuration of M1 in the $YZ$ plane is not perfectly conserved, so it slightly tilts in the $X$ direction and moves out of the $YZ$ plane. This distortion turns out to be much smaller for more symmetric M6 with double Ni$-$O bonds on both sides.  
As a result, the HOMO of M1 is expected to overlap more with Ni apex $s$-states which lead to broader spin up transmission peak compared to M6 case. A similar result was also reported previously in nonmagnetic molecular junctions which can switch between high and low conductance states by a mechanical strain \cite{Quek-2009}. The second reason is the rather strong localization of HOMO and HOMO-1 of M6 on the linking groups (the orbitals are decoupled in the middle) compared to a rather delocalized HOMO of M1 (see Fig. \ref{homo-lumo}a,f). Concerning the stability of molecular junctions with NO$_2$$-$terminations, we note that two contradictory results have been reported. Zotti $et$ $al$ \cite{zotti2010revealing} concluded that NO$_2$$-$terminated tolanes form rather stable molecular junctions under ambient conditions with MCBJ while Kaliginedi $et$ $al$ \cite{Kaliginedi2014} showed that molecular junctions formed with NO$_2$$-$caped molecule are rather unstable. The authors argued that this difference may arise from different experimental conditions in two studies. Moreover, Vardimon $et$ $al$ \cite{Vardimon2015} successfully created nickel oxide atomic junctions due to presumably strong chemisorption at the Ni$-$O contacts. Therefore, we hope that the M6 ($-$NO$_2$) junction could be stable with Ni electrodes which needs, of course, to be verified by future experiments.

\begin{figure}[htbp]
	\centering
	\includegraphics[scale=0.55]{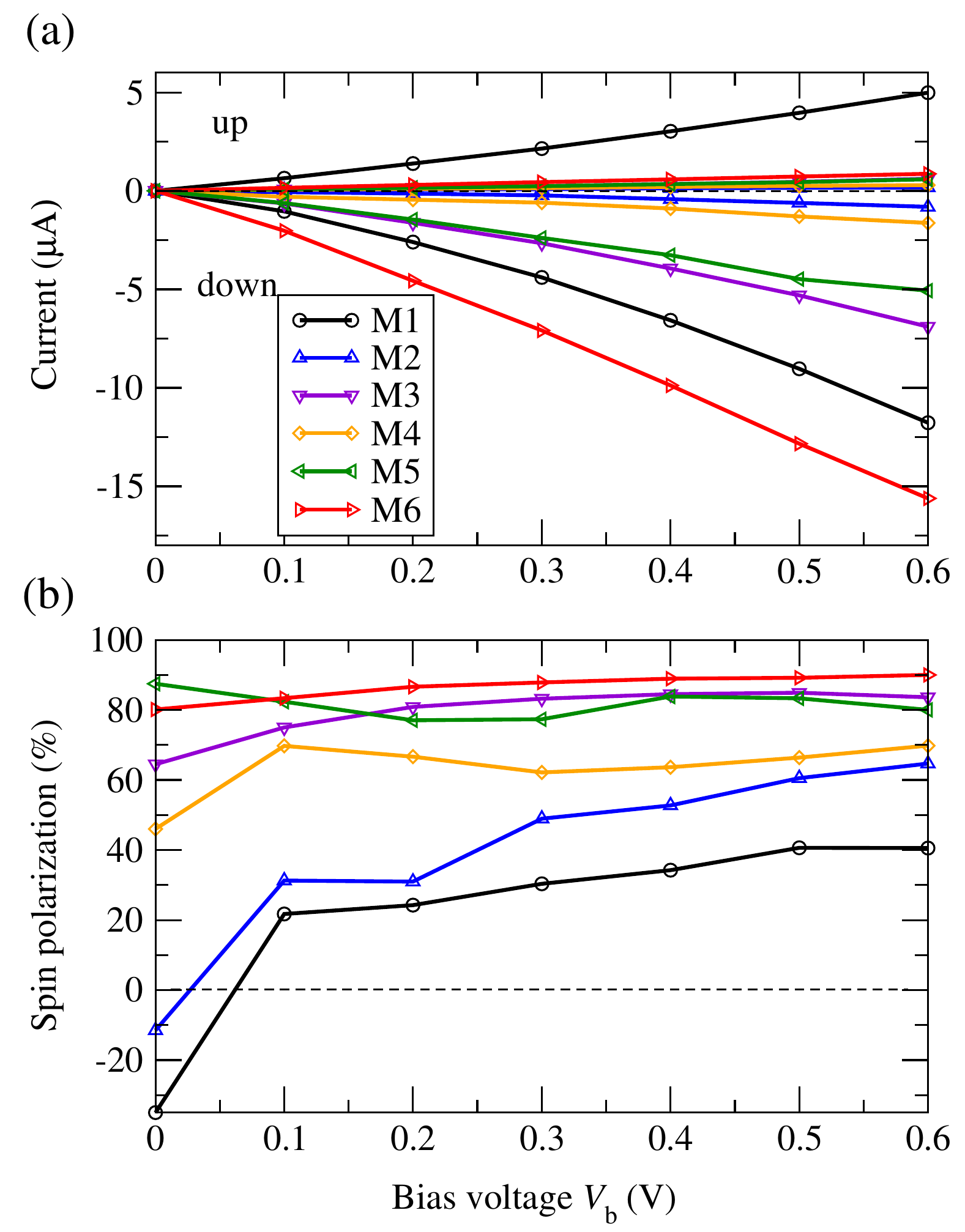}
	\caption{\label{polarization} (a) Current-voltage characteristics for spin up (positive values) and spin down (negative values) channel for M1, M2, M3, M4, M5 and M6 molecular junctions, (b) Corresponding SP as a function of voltage. Note that conductance values are used to evaluate SP in equilibrium case. }
\end{figure}

We now turn our attention to the electron-accepting groups, {\it i.e.}, M2, M3, M4, and M5. Here, the transport is dominated by LUMO. For M2, when a thiolate ($-$S) was replaced by a methylthiol ($-$CH$_3$S), the molecule-metal coupling strength is significantly reduced, resulting in a narrower LUMO resonance peak at about 1.5 eV. Moreover, due to LUMO symmetry (Fig. \ref{homo-lumo}b) it does not overlap with Ni apex $d^\downarrow_{xz,yz}$ states which explains that no increase of spin down transmission is observed at around $0.4$ eV where those $d$-states dominate the Ni down DOS. That explain rather low spin down conductance and SP for M2 compared to M1 case. Our results have a general agreement with very recent experimental measurements on thiolate and methylthiol terminated systems \cite{Leary2018,inkpen2019non} with Au electrodes. Moreover, in Ref. \cite{inkpen2019non}, the authors confirmed experimentally that S$-$Au and CH$_3$S$-$Au are chemisorption and physisorption mechanisms, respectively. Next, if we replace one of $-$CH$_3$ by $-$COOH and $-$CNH$_2$NH, forming M3 and M4 asymmetric junctions, the LUMO approaches to $E_F$, 
leading to enhanced conductance. For these molecules, the LUMO (Fig. \ref{homo-lumo}c,d) will overlap now with $d^\downarrow_{xz,yz}$ Ni states 
so that larger spin down conductance (and noticeable SP)  is again recovered for M3 and M4 junctions which show in fact rather similar transmission features. In the case of M3 junction, it has been shown experimentally that the formations of $-$COO$^-$ and $-$COOH in solution depends on pH conditions \cite{Chen-2006,Ahn-2012}. Note that Sheng $et$ $al$ have also considered $-$COOH in their $ab$ $initio$ calculations \cite{Sheng2009} of alkane molecular wires. 
We present therefore in Fig. \ref{T-E-curve}c the results for COOH$-$Ni contact, as shown in Fig. \ref{structure}c. In addition, we also investigated the M3 junction with one removed ``H" atom forming COO$^-$$-$Ni bond at the interface (see Appendix D). Interestingly, as shown in Fig. \ref{Sup-fig3}, when the ``H" atom is removed, the charge transport is dominated by HOMO rather than LUMO due to lack of one electron compared to the neutral molecule. Moreover, the conductance values for both spins are found one order of magnitude smaller than for neutral molecule due to very sharp HOMO and its localization around ``O" atoms at the interface (see inset in Fig. \ref{Sup-fig3}b). Finally, for M5 molecule rather broad LUMO-derived transmission feature is seen about 0.75 eV, significantly increasing for spin down channel upon approaching the Fermi level due to noticeable coupling to Ni spin down $d$-states. We note that the DFT transport scheme tends to overestimate the conductance of molecular junctions relative to experiments due to underestimation of the HOMO-LUMO gap. $GW$ self-energy has been recognized as a good approach to correct for this error and to describe more accurately the energy level alignment at the molecule-metal interface. However, it has been shown that $GW$ corrections affect rather unoccupied orbitals (poorly described within the ground state DFT) while occupied levels (in particular, the HOMO) are only slightly altered \cite{Strange-2011, Rangel2017}. Since the charge transport in M1 and M6 is dominated by the HOMO level, we hope that our mean-field DFT studies are good enough to provide reliable comparative results. On the other hand, for LUMO dominant junctions such as M2, M3, M4 and M5, smaller conductance values are found in our calculations (for both spin channels) which are expected to be even smaller when more sophisticated $GW$ self-energy approach is used. Moreover, the DFT error is in general systematic, thus conductance ratios are usually in good agreement with experiment, for instance, the rectification ratios predicted by NEGF-DFT were found to be reliable \cite{elbing2005single, Batra2013}. Therefore, we believe that the spin filtering ratios presented in this work are reliable as well and that the DFT-error introduced here plays a minor role in a quantitative basis but should not affect our main conclusions.

Having understood the transport properties at equilibrium, we now turn to the out-of-equilibrium situation with a small bias voltage ($V_{\text{b}}$), up to 0.6 V. At each voltage, the spin up and down currents are determined self-consistently under non-equilibrium condition using the Landauer-B\"{u}ttiker formula (see Eq. \ref{Landauer}). The results show various trends for different anchoring groups. Clearly, spin down current (plotted with negative values) is significantly larger than the corresponding spin up one (plotted with positive values) as shown in Fig. \ref{polarization}a. More importantly, for M6, the spin up current increases very slowly with an approximately linear trend. On the contrary, the spin down current increases much more rapidly which results in a high spin injection efficiency. The pronounced increase of spin down currents for M1 and M6 indicates a strong metal/molecule orbital coupling close to $E_F$. On the contrary, for M3 and M5, the current for both spin channels increase much slower. For spin up channel, small currents are attributed to the fact that the LUMO resonance lies at about 0.5 and 0.7 eV (see Fig. \ref{T-E-curve}) for M3 and M5, respectively, which are not included in the explored bias window. In addition, we find that M2 and M4 has the lowest currents for both spin channels.

\begin{figure}[htbp]
	\centering
	\includegraphics[scale=0.46]{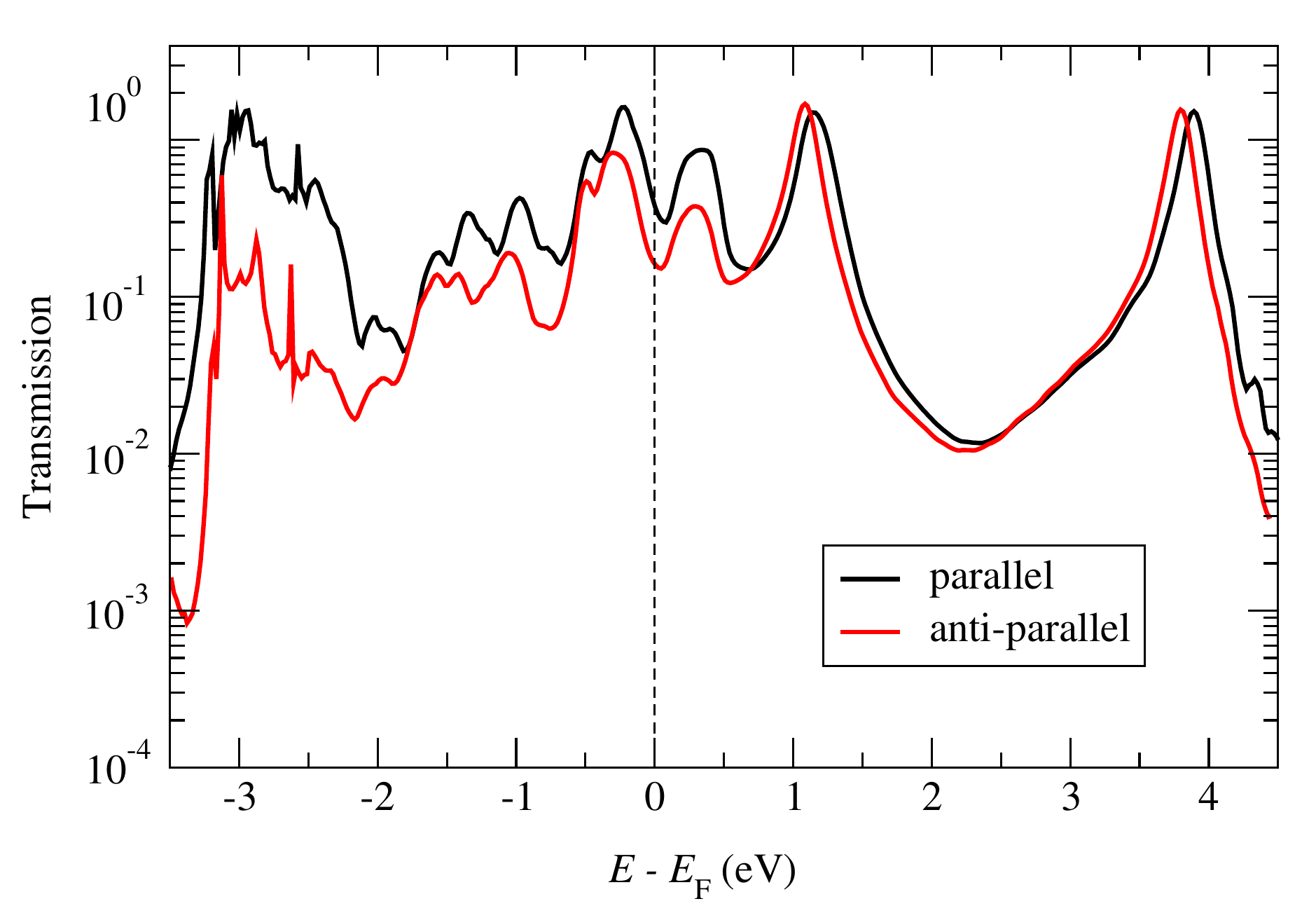}
	\caption{\label{AP-M6} Total transmission functions (in logarithmic scale) of M6 junction at the zero-bias voltage for both parallel and anti-parallel magnetic configurations of two Ni electrodes. Note that spin up and down contributions (equal due to symmetry in the antiparallel case) are summed in both configurations. The calculated MR was found to be as large as 140\%.}
\end{figure}

The spin polarization of the current at a bias voltage is evaluated as, SP = $(I_{\downarrow}-I_{\uparrow})/(I_{\uparrow}+I_{\downarrow}) \times 100\%$, where $I_{\uparrow}$ and $I_{\downarrow}$ are spin up and down currents, respectively. Note that at equilibrium, the SP was evaluated from conductance values. For zero bias voltage, the SP for M5 and M6 are more than 80$\%$. Interestingly, when the bias voltage is applied, the SP of M6 is further enhanced up to more than 90\%. On the contrary, the SP of M5 is slightly decreased. Moreover, for M3, the SP is less than 65$\%$ under zero bias, but it can become as large as 80$\%$ at $V_{\text{b}} = 0.6$ V. On the other hand, the SP of M4 slightly increases when the bias voltage is applied. Interestingly, the SP of M1 and M2 changes sign when the bias voltage is applied but remains relatively low.

Since the M6 exhibits large SP as well as high conductance, we have also studied its magnetoresistance (MR) property which measures the change in total current between the parallel (P) and antiparallel (AP) magnetic alignments of Ni electrodes. 
In the linear regime it can can be calculated as MR = $(G_{\text{P}}-G_{\text{AP}})/G_{\text{AP}} \times 100\%$, where $G_{\text{P}}$ and $G_{\text{AP}}$ are total conductances (sum of spin up and down contributions) at zero bias for P and AP magnetic configurations, respectively. We present in Fig. \ref{AP-M6} the total transmissions for two magnetic cases. As expected, in addition to the large SP the M6 junction also shows very high MR, as large as 140\%. It is much larger than the MR found for M1 junctions, of about 27\%, as reported in Ref.\cite{Waldron-2006}, confirming once again very good spin filtering properties of symmetric NO$_2$ anchoring groups. In Appendix D, we summarized the transmission functions of anti-parallel spin configurations (Fig. \ref{Sup-fig4}) and corresponding MR (Table 
\ref{table-2}) values for all six molecular junctions.

\section{ Conclusions}
\label{conclusions}

To conclude, by using a combination of density
functional theory and non-equilibrium Green's function formalism, we have investigated the effect of anchoring groups on spin-polarized transport through benzene-derivative molecular junctions joining two ferromagnetic Ni(111) electrodes. It was found that anchoring groups have a strong impact on the energy alignment of relevant molecular orbitals with respect to the Fermi level and the degree of molecule-metal hybridization. Therefore, the choice of anchoring groups indeed strongly affects the conductance of the molecular junction and its spin polarization, SP. According to our study, M6 ($-$NO$_2$) junctions exhibit overall the best performance with high conductance (and also the current), large SP ($>$80\%) as well as giant MR of about 140\%. Interestingly, the SP can be further enhanced (up to 90\%) by a small voltage. It was attributed to a rather sharp/broad HOMO-derived resonance in spin up/down transmission around the Fermi energy dictated by the HOMO symmetry and its spatial distribution. The S and CH$_3$S systems, on the contrary, exhibit rather low SP while intermediate values are found for COOH and CNH$_2$NH groups. It has been found, in addition, that the large SP of M5 ($-$NC) is slightly decreased with the voltage. We believe that our comparative and systematic studies will enrich the understanding of the role of anchoring groups on spin-polarized transport of molecular junctions and will be useful for further studies and applications in molecular spintronics.

--------------------------------------------
\section{Acknowledgments}
D.L. was supported by the Alexander von Humboldt Foundation through a Fellowship for Postdoctoral Researchers. 
\\

\section*{Appendix A: Comparison between $\textsc{SIESTA}$ and $\textsc{QE}$ results}

\begin{figure*}[htbp]
	\centering
	\includegraphics[scale=0.45]{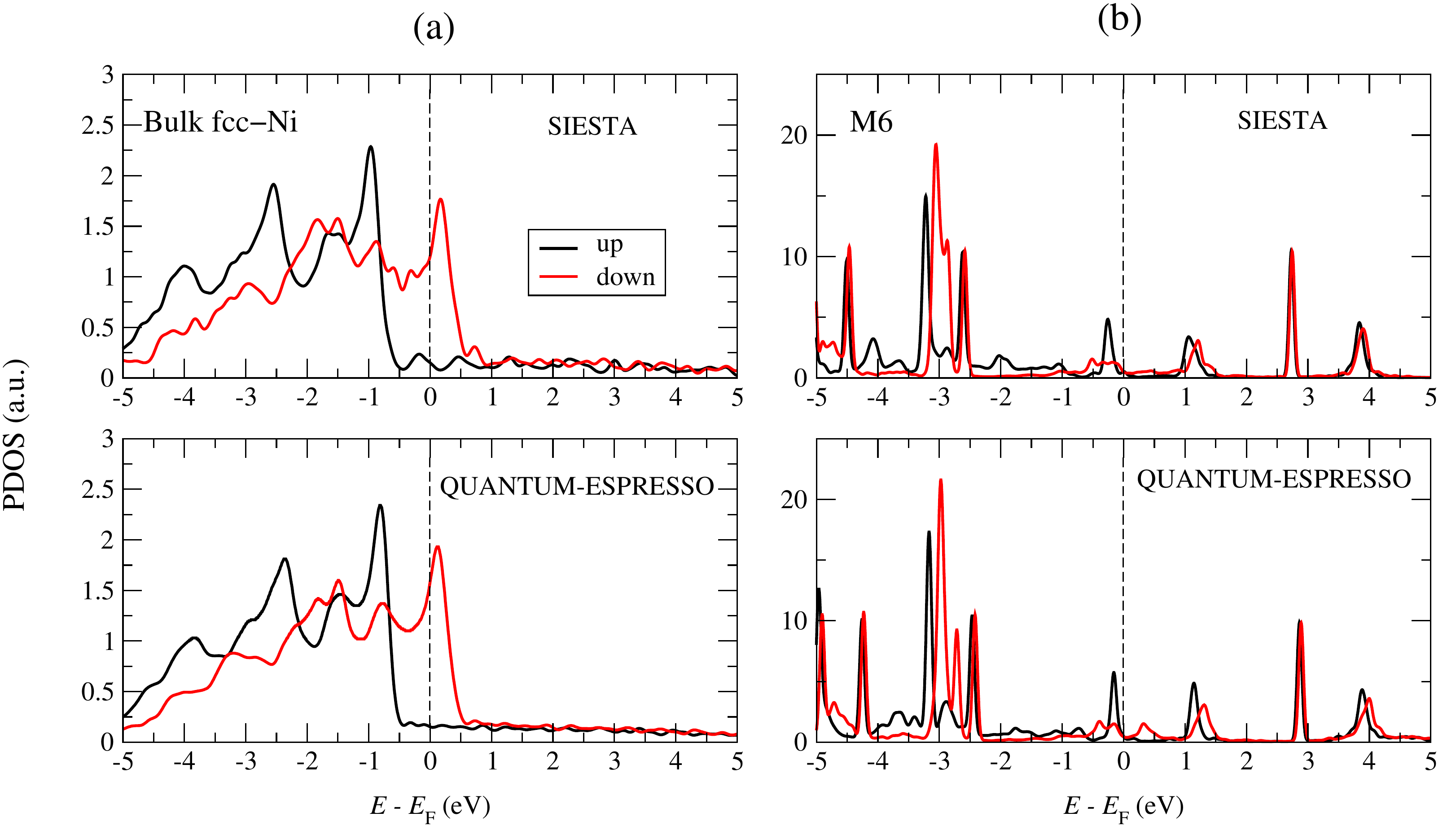}
	\caption{\label{Sup-fig1}
		(a) Spin-resolved total density of states (DOS) of bulk fcc-Ni for spin up (black) and down (red), calculated by $\textsc{siesta}$ (top) with NCPP and by $\textsc{QE}$ with ultrasoft PP (down). The exchange splitting from $\textsc{siesta}$ is about 0.16 eV larger, which results in a 0.05 $\mu_{\text{B}}$ larger magnetic spin moment compare to QE results. (b) Spin-dependent projected DOS on molecule for M6 junction calculated by $\textsc{siesta}$ and $\textsc{QE}$. A very good overall agreement is found between $\textsc{siesta}$ and $\textsc{QE}$ results.
	}
\end{figure*}

For fcc-Ni as seen in Fig. \ref{Sup-fig1} (a), the magnetic moments calculated by $\textsc{QE}$ was found to be about 0.65 $\mu_{\text{B}}$ while the $\textsc{siesta}$ within DZP basis gives about 0.70 $\mu_{\text{B}}$. The small discrepancy between QE and $\textsc{siesta}$ on spin moment can be traced to the use of NCPPs in $\textsc{siesta}$, versus ultrasoft PPs in QE. The similar results were also reported in Ref.\cite{cakir-2014, Pablo-2015}. In addition, the single $\zeta$ (SZ) and SZP basis sets of $\textsc{siesta}$ give the spin moment of 0.78 $\mu_{\text{B}}$ and 0.74 $\mu_{\text{B}}$, respectively, suggesting that the DZP basis set for Ni is the best one in terms of a more accurate description of spin moment.

In order to check the reliability of our DZP basis sets used in this work, we also compared the spin-dependent projected DOS on the molecule for M6 junctions as plotted in Fig. \ref{Sup-fig1}b. A good agreement between QE and $\textsc{siesta}$ results is found in terms of energy level alignments, indicating the validity of our DZP basis set.  

\section*{Appendix B: Projected DOS on molecules with the parallel spin configuration}

\begin{figure*}[htbp]
	\centering
	\includegraphics[scale=0.6]{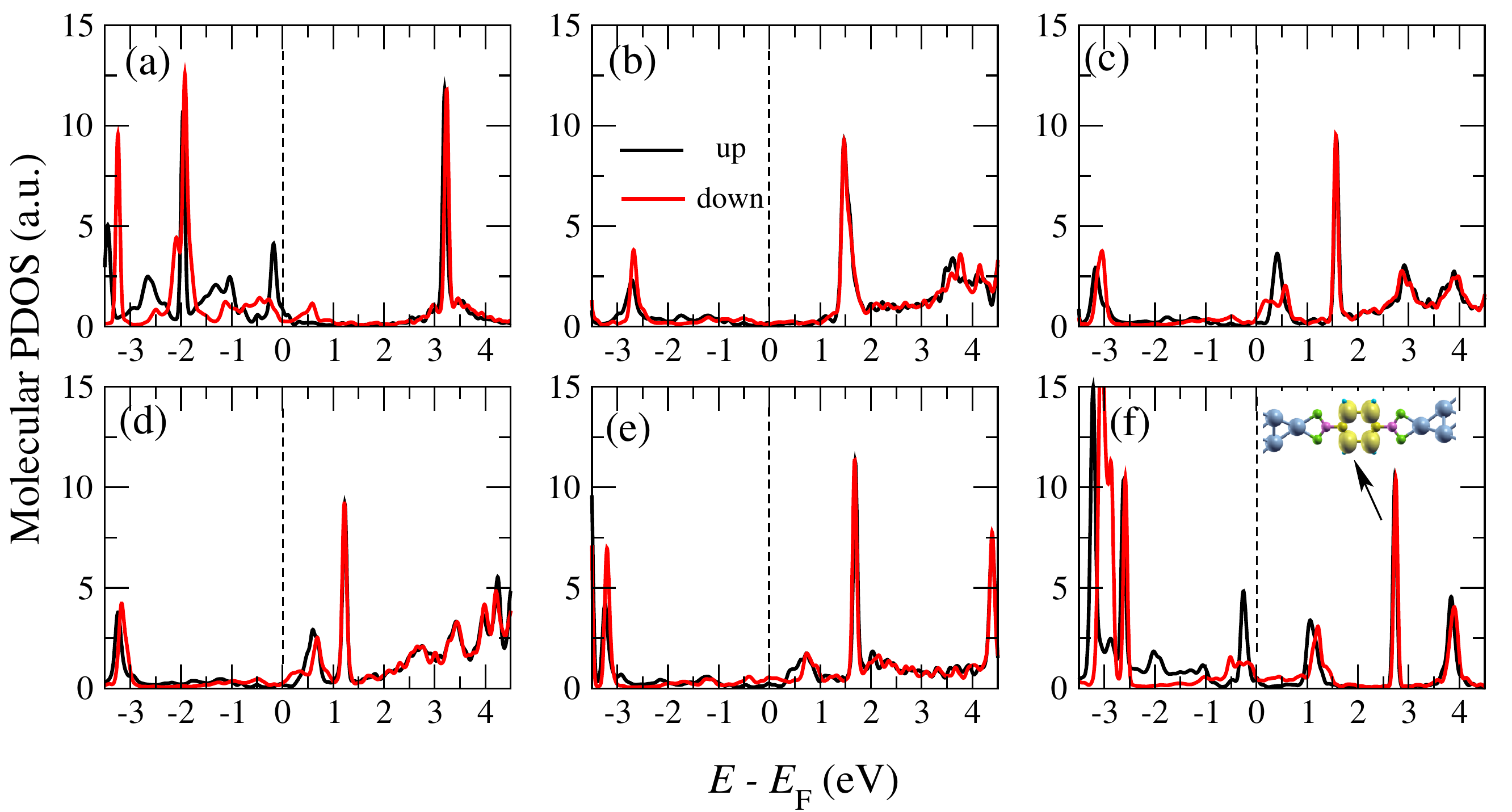}
	\caption{\label{Sup-fig2}
		Spin-resolved projected DOS on the molecule with the parallel magnetic alignment of two Ni electrodes for (a) M1, (b) M2, (c) M3, (d) M4, (e) M5 and (f) M6 molecular junctions. Spin up and down channels are plotted by black and red lines, respectively. The local DOS in the energy window of LUMO+1 for M6 is also plotted in the inset. Note that the zero of energy is at the Fermi level. 
	}
\end{figure*}

To identify the positions of molecular levels, we display in Fig. \ref{Sup-fig2} the projected DOS on the molecule with the parallel magnetic alignment of two Ni electrodes for six molecular junctions. HOMO and LUMO peaks are clearly seen.

Interestingly, we found that LUMO+1 of M6 (see inset of Fig. \ref{Sup-fig2}f) is strongly localized on 4 carbon atoms of the molecule and is completely decoupled from electrodes, which explains that no LUMO+1 derived peak was observed in the transmission curve plotted in Fig. \ref{T-E-curve} (f).

\section*{Appendix C: M3 junction with removed single ``H"}

\begin{figure*}[htbp]
	\centering
	\includegraphics[scale=0.5]{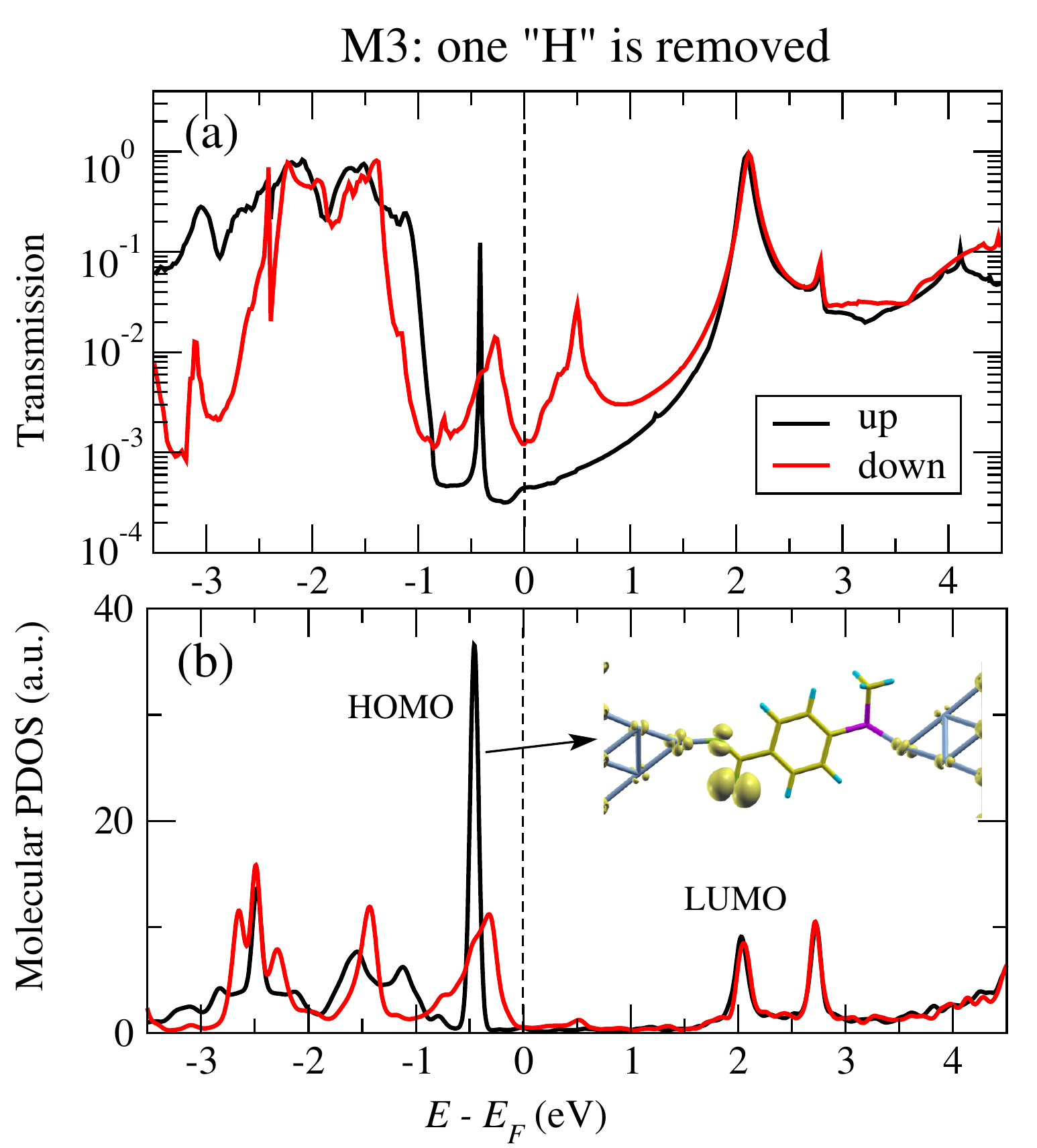}
	\caption{\label{Sup-fig3}
		M3 molecular junction with COO$^-$$-$Ni contact, here one ``H" atom is removed. Spin-resolved transmission function (a) and projected DOS on the molecule (b). Spin up and down channels are plotted by black and red lines, respectively. Note that the zero of energy is at the Fermi level. Local DOS at the energy range of HOMO peak for spin up is plotted in the inset.
	}
\end{figure*}

When the ``H" atom is removed, the electron transport is dominated by HOMO rather than LUMO due to loss of one electron (see Fig. \ref{Sup-fig3}). Interestingly, the conductance values for both spins are bout one order of magnitude smaller than corresponding M3 junction because of super sharp HOMO and its localized features around ``O" atoms at the interface (see inset in Fig. \ref{Sup-fig3}b).

\section*{Appendix D: Spin-dependent $\textit{T(E)}$ with the anti-parallel spin configuration}

\begin{figure*}[htbp]
			\centering
			\includegraphics[scale=0.60]{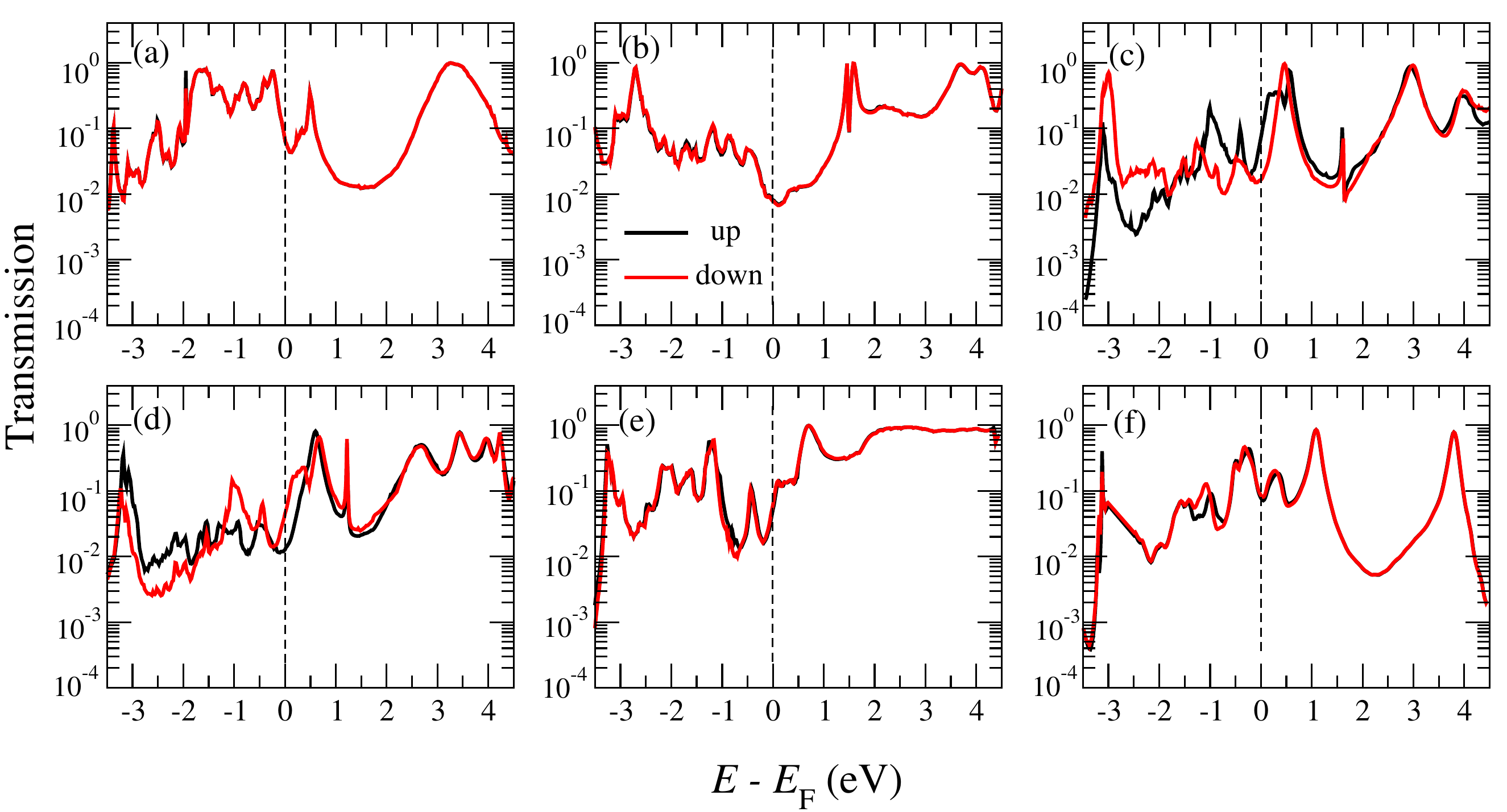}
			\caption{\label{Sup-fig4}
				Spin-resolved zero-bias transmission functions (in logarithmic scale) with the anti-parallel magnetic alignment of two Ni electrodes for (a) M1, (b) M2, (c) M3, (d) M4, (e) M5 and (f) M6 molecular junctions. Spin up and down channels are plotted by black and red lines, respectively. Note that the zero of energy is at the Fermi level. 
			}
\end{figure*}

We present in Fig. \ref{Sup-fig4} the spin-dependent transmission functions for the anti-parallel magnetic configuration of two Ni electrodes. Due to symmetry, spin up and spin down $T(E)$ superpose for symmetric junctions (M1, M2, M5, and M6) while they are slightly different for asymmetric cases (M3 and M4). The corresponding magnetoresistance values are summarized in Table. \ref{table-2}.

\begin{table}[h]
	\centering
	\scalebox{1.0}{
		\begin{tabular}{ccccccc}
			\hline\hline
			\multicolumn{1}{c}{$$} & \multicolumn{1}{c} {M1} & \multicolumn{1}{c} {M2} & \multicolumn{1}{c} {M3} & \multicolumn{1}{c} {M4} & \multicolumn{1}{c} {M5} & \multicolumn{1}{c} {M6}\\ \hline
			MR (\%) & 35  & 10 & 30 & 38 & 92 & 140 \\ \hline
			\end{tabular}}
		\caption{Calculated magnetoresistance, MR = $(G_{\text{P}}-G_{\text{AP}})/G_{\text{AP}} \times 100\%$, of six molecular junctions.} \label{table-2}
\end{table}

\bibliographystyle{apsrev}
\bibliography{References}

\end{document}